\documentclass{ieeeaccess}
\usepackage{cite}
\usepackage{amsmath,amssymb,amsfonts}
\usepackage{algorithmic}
\usepackage{graphicx}
\usepackage{textcomp}
\usepackage{booktabs, makecell}
\def\BibTeX{{\rm B\kern-.05em{\sc i\kern-.025em b}\kern-.08em
    T\kern-.1667em\lower.7ex\hbox{E}\kern-.125emX}}

\begin{document}
\doi{}

\title{Automated Huntington's Disease Prognosis via Biomedical Signals and Shallow Machine Learning}
\author{\uppercase{Sucheer Maddury}\authorrefmark{1}}
\address[1]{Leland High School, San Jose, California 95120, USA (e-mail: sumaddurycollege2024@gmail.com)}


\corresp{Corresponding author: Sucheer Maddury (e-mail: sumaddurycollege2024@gmail.com).}

\begin{abstract}
Huntington's disease (HD) is a rare, genetically-determined brain disorder that limits the life of the patient, although early prognosis of HD can substantially improve the patient's quality of life. Current HD prognosis methods include using a variety of complex biomarkers such as clinical and imaging factors, however these methods have many shortfalls, such as their resource demand and failure to distinguish symptomatic and asymptomatic patients. Quantitative biomedical signaling has been used for diagnosis of other neurological disorders such as schizophrenia, and has potential for exposing abnormalities in HD patients. In this project, we used a premade, certified dataset collected at a clinic with 27 HD positive patients, 36 controls, and 6 unknowns with electroencephalography, electrocardiography, and functional near-infrared spectroscopy data. We first preprocessed the data and extracted a variety of features from both the transformed and raw signals, after which we applied a plethora of shallow machine learning techniques. We found the highest accuracy was achieved by a scaled-out Extremely Randomized Trees algorithm, with area under the curve of the receiver operator characteristic of 0.963 and accuracy of 91.353$\%$. The subsequent feature analysis showed that 60.865$\%$ of the features had p$<$0.05, with the features from the raw signal being most significant. The results indicate the promise of neural and cardiac signals for marking abnormalities in HD, as well as evaluating the progression of the disease in patients.
\end{abstract}


\titlepgskip=-15pt

\maketitle

\section{Introduction}
\label{sec:introduction}
\hspace{0.5cm}Huntington's disease (HD) is a neurodegenerative disease characterized by motor, cognitive, and/or behavioural disturbances, which manifest themselves in symptoms at a mean onset of 30-50 years of age \cite{HuntingtonDisease}. HD is autosomal dominant, and the genetic cause is repeats in cytosine-adenine-guanine (CAG) repeats in the Huntingtin (HTT) gene, with a clear inverse relationship between the number of CAG trinucleotide repeats and the age of HD onset \cite{HuntingtonDisease}. The CAG repeats causes mutational expansion of polyglutamine in HTT proteins, resulting in protein folding restriction, with prevalence of the mutation varying widely between geographical regions with an average worldwide prevalence of 2.7 HD positive per 100,000 \cite{HuntingtonDisease,HuntingtonPrevalance}. The disease is most prevalent in Western populations, however, at a rate of 10.6-13.7 HD-positive individuals per 100,000 people--with almost 30,000 HD patients in the United States \cite{HuntingtonClinicals}.

\hspace{0.5cm}HD currently has no cure, although new clinical approaches such as Antisense oligonucleotide therapy (AOT) have shown some promise for reducing levels of the mutation \cite{HuntingtonClinicals}. Early detection of HD, however, is crucial for clinical trials and intervention in the earliest stages of the disease, which can benefit the quality of life of the patients \cite{EarlyDetection}. Current diagnosis techniques rely on a combination of behavioural, clinical, imaging, and genetic factors as well as a complex analysis into the HTT gene in the family. Such approaches have several drawbacks, including being time and resource-costly, as well as failing to distinguish between asymptomatic and symptomatic carriers of the HTT mutation \cite{Currentdiagnosis}. Many symptoms that mimic HD also exist, presenting a prognostic challenge in certain patients, which can be exacerbated in poorer or rural hospitals \cite{Currentdiagnosis}. The relative rarity of the disease has also contributed to a shortfall in understanding of HD at a very specific, symptomatic level, signaling the need for more robust diagnostic solutions.

\hspace{0.5cm}Several past studies have examined HD diagnosis in detail. Paulsen et al. (2008) studied hundreds of HD gene-positive patients in an attempt to predict Huntington's progression decades in advance \cite{BackgroundOne}. Each patient underwent several lengthy trials measuring genetic, neurobiological, clinical, and behavioural factors through blood tests, neurological and cognitive exams, psychological questionnaires, brain imaging, and more. The study found that the time of HD diagnosis could best be predicted by clinical and neuroimaging markers (p$<$0.0001). Paulsen (2009) examined past literature on MRI and fMRI for Huntington's disease prognosis \cite{BackgroundThree}. While there are limitations, the study describes the potential for magnetic resonance imaging for detective sensitive changes resulting from the earliest progressions of HD. Lastly, Mason et al. (2018) also focused on neuroimaging markers for early diagnosis of HD, with the objective of correcting for the imprecisity of statistical models based on motor symptoms \cite{BackgroundTwo}. Through resting-state and structural fMRI, the researchers trained a support-vector machine on a cohort of 19 HD-positive patients and 21 controls. The researchers found that a holistic imaging approaches resulted in high classification accuracy (p$<$0.03).

\hspace{0.5cm}There is also literature on the altered electrical activity throughout the body in Huntington's patients, particularly in electroencephalography (EEG) and electrocardiography (ECG), both tests that measure electrical activity in various regions through sensitive electrodes. Scott et al. (1972) measured EEG levels at all voltages and frequencies in the brain in HD-positive and control patients \cite{EEGThree}. They found that the presence of low voltage EEG, particularly in the delta and theta frequency bands, specified cortical atrophy often found in HD, while low voltage EEG was rarely present in control patients. Cankar et al. (2018) tested ECG in HD-positive and healthy control patients, since one of the most common causes of death in HD patients is heart disease \cite{EEGFour}. The researchers found that HD-positive individuals exhibit higher voltage variability in the ECG regions among standard QT intervals (p$<$0.01), signifying cardio-electric remodeling in HD patients. 

\hspace{0.5cm}Biological signaling can be carried out quickly at low cost, and quantitative EEG (qEEG) has already shown promise for early detection of other neurodegenerative diseases such as Alzheimer's disease and Parkinson's disease \cite{EEGOne,EEGTwo}. In this study, one of the first of its kind, we will test classical machine learning trained on EEG, ECG, and functional near-infrared spectrography (fNIRS). Then, the model will be optimized and evaluated, with the goal of assessing the potential for neural and cardiac signaling for HD prognosis.

\section{Materials and Methods}
\label{sec:materialsandmethods}
\subsection{Dataset Overview}

\hspace{0.5cm}Data for this study came from an HD and controls dataset published publicly by Lancaster University \cite{data}. In the data collection process, 69 participants were recruited from the Neurological Clinic in Ljubljana, Slovenia to undergo several tests. As seen in Table~\ref{tab:tab1}, of these participants, 15 were symptomatic HD (SHD), 12 were pre-symptomatic HD (PHD), 36 were health controls and 6 were unclassified. Preliminary testing showed that the unclassified patients had extremely similar features to the control patients, and were thus considered controls for the remainder of this study. Over the course of 30 minutes (then cut to 20 minutes), EEG, ECG,  fNIRS, respiration, and blood flow data was recorded via a variety of instruments, and the room and patient forehead temperature was also taken during this time. The EEG signal was recorded through V-Amp headset with 16 electrodes (10-20 system) at 1 kHz, with the participant sitting in a chair with eyes open. The specific electrodes used were the C3, C4, Cz, F3, F4, Fp1, Fp2, O1, O2, P3, P4, P7, P8, Pz, T7, T8 placements. The ECG signal was recorded over a single channel at 1.2 kHz with sensors placed on the shoulders and lower left rib. The fNIRS data was taken with 11 optodes (10-20 system) at 31.25Hz in placements N1, N2, N3, N4, N5, N6, N7, N8, N9, N10, and N11, each with deoxygenated and oxygenated channels (giving 22 channels in total). The data collection protocols in the dataset used in this study were approved by the commission of the Republic of Slovenia for Medical Ethics.
\begin{table*}[]
\centering
\caption{Dataset signal collection information.}
\label{tab:tab1}
\resizebox{6in}{!}
{
\begin{tabular}{@{}llllllll@{}}
\toprule
\textbf{}          & \textbf{}           & \textbf{EEG}             & \textbf{}               & \textbf{ECG}                    &                         & \textbf{fNIRS}        &                         \\ \midrule
\textbf{Diagnosis} & \textbf{Count (\#)} & \textbf{Electrodes (\#)} & \textbf{Frequency (Hz)} & \textbf{Optodes (\#)} & \textbf{Frequency (Hz)} & \textbf{Optodes (\#)} & \textbf{Frequency (Hz)} \\
SHD                & 15                  & 16                       & 1000                    & 11                             & 1200                    & 11                    & 31.25                   \\
PHD                & 12                  & 16                       & 1000                    & 11                             & 1200                    & 11                    & 31.25                   \\
Control            & 36                  & 16                       & 1000                    & 11                             & 1200                    & 11                    & 31.25                   \\
Unknown            & 6                   & 16                       & 1000                    & 11                             & 1200                    & 11                    & 31.25                   \\ \bottomrule
\end{tabular}%
}
\end{table*}
\subsection{EEG, ECG, and fNIRS Signal Processing}
\hspace{0.5cm}To be used for HD classification, the EEG, ECG, and fNIRS signals had to be preprocessed and features extracted, performed mostly using MNE (https://mne.tools/) \cite{MNE}. As seen in Figure~\ref{fig:fig1}, signal preprocessing was divided into formatting/cutting, filtering, segmentation, and finally feature extraction. First, the EEG, ECG, and fNIRS data were separated from the rest of the data and concatenated into their own respective CSV files from the original MAT files.
\Figure[t!][width=\linewidth]{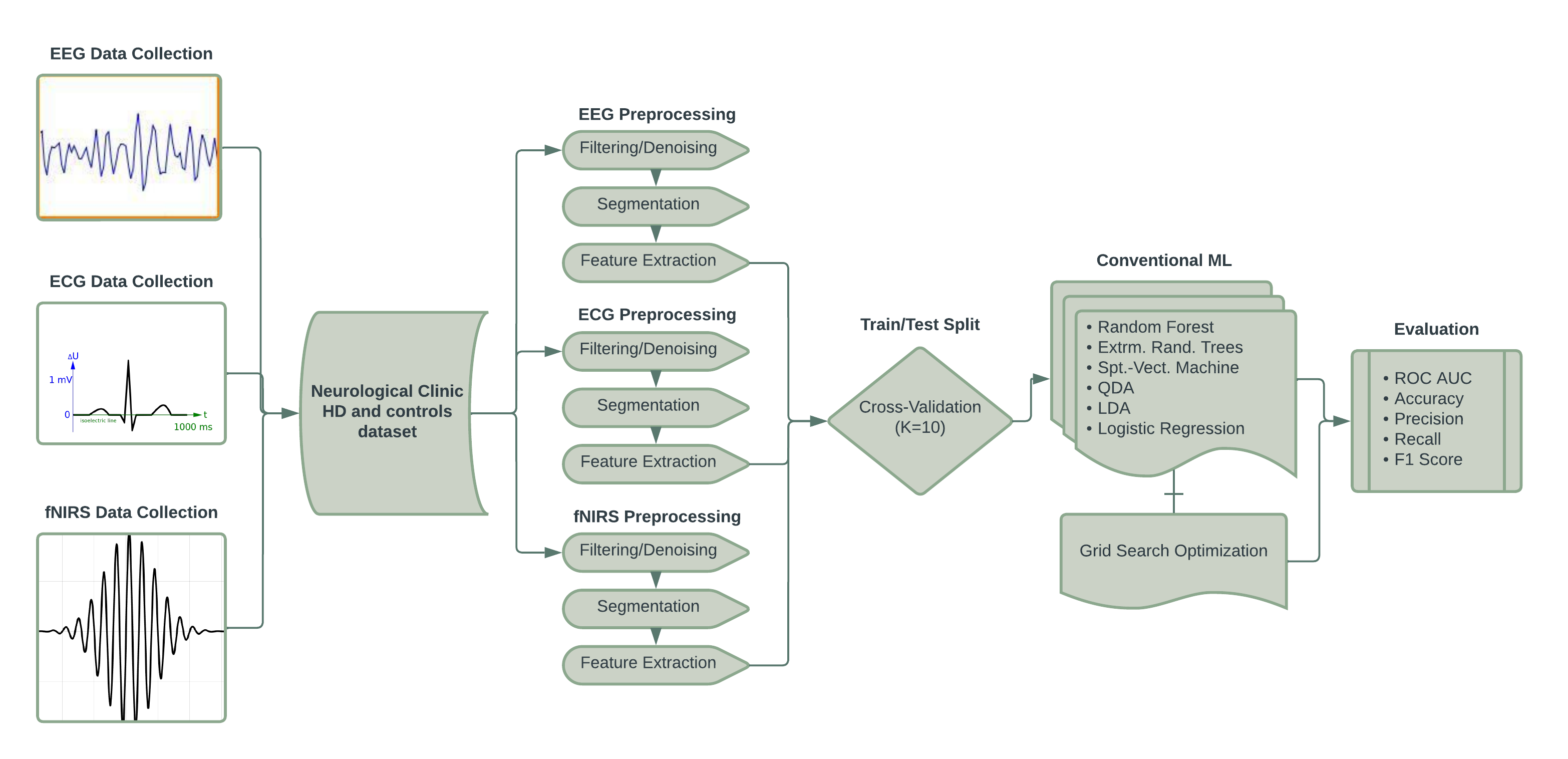}
{Biological signal data applications flowchart.\label{fig:fig1}}

\begin{figure*}
(A)\includegraphics[width = 3in]{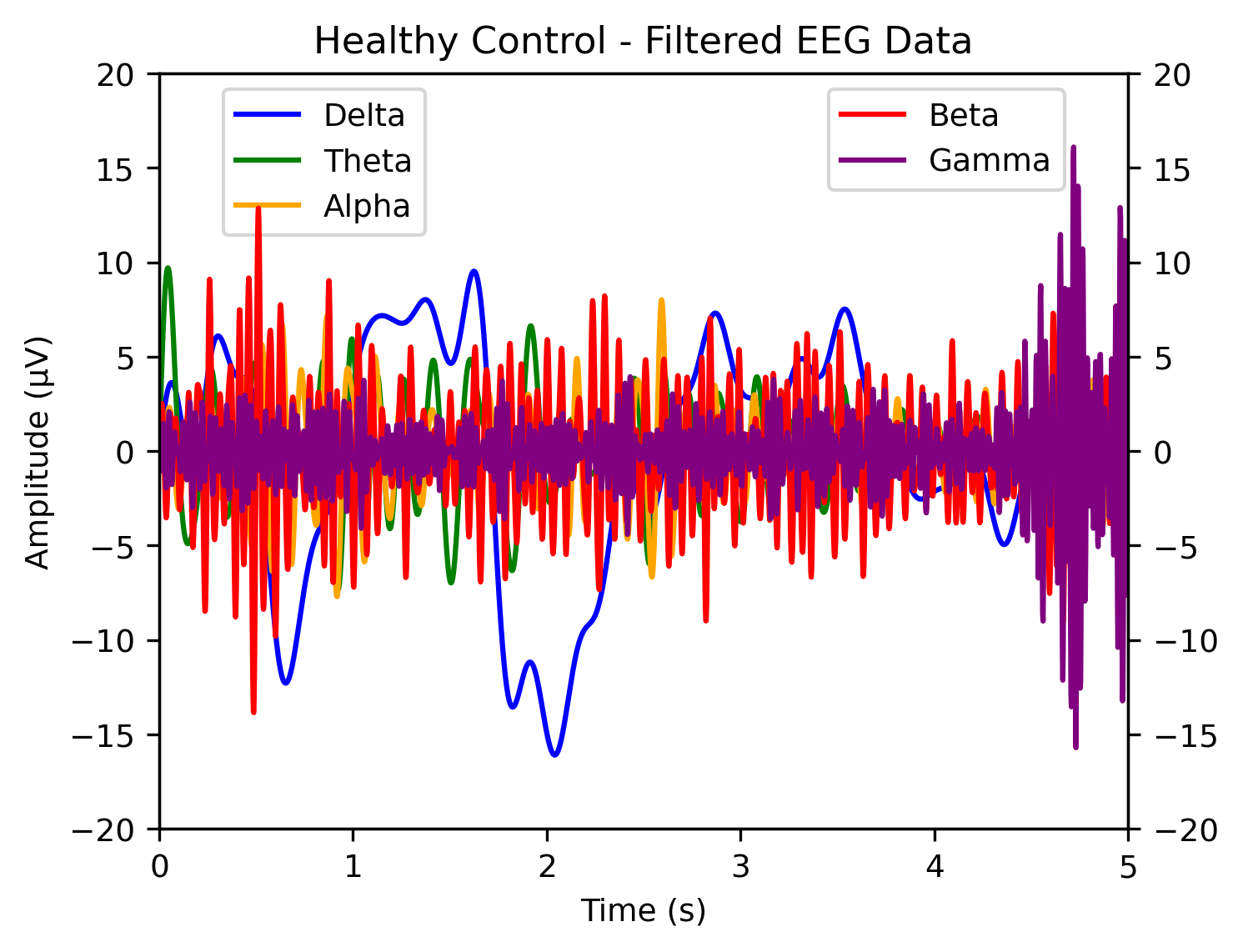}
(B)\includegraphics[width = 3in]{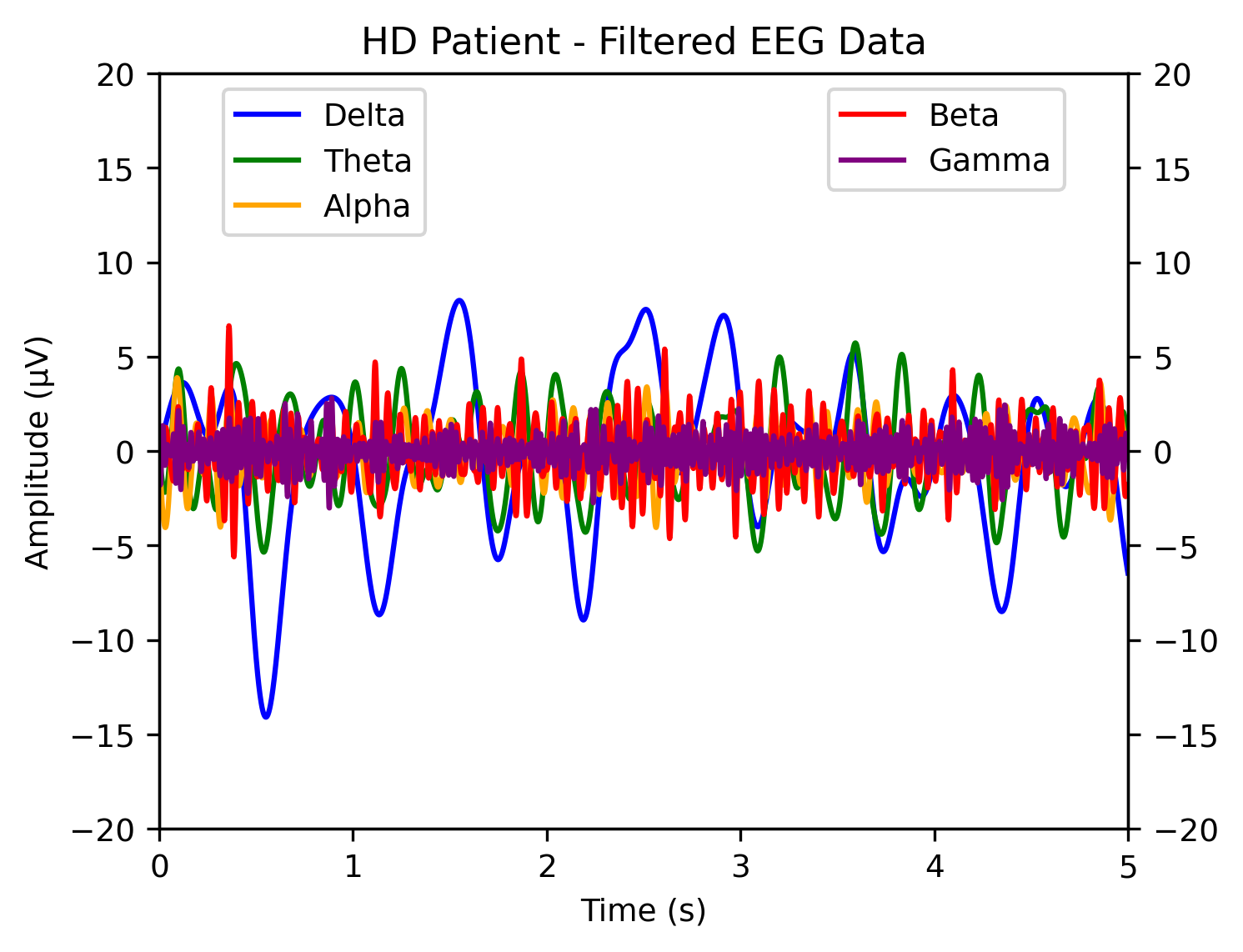}\\
(C)\includegraphics[width = 3in]{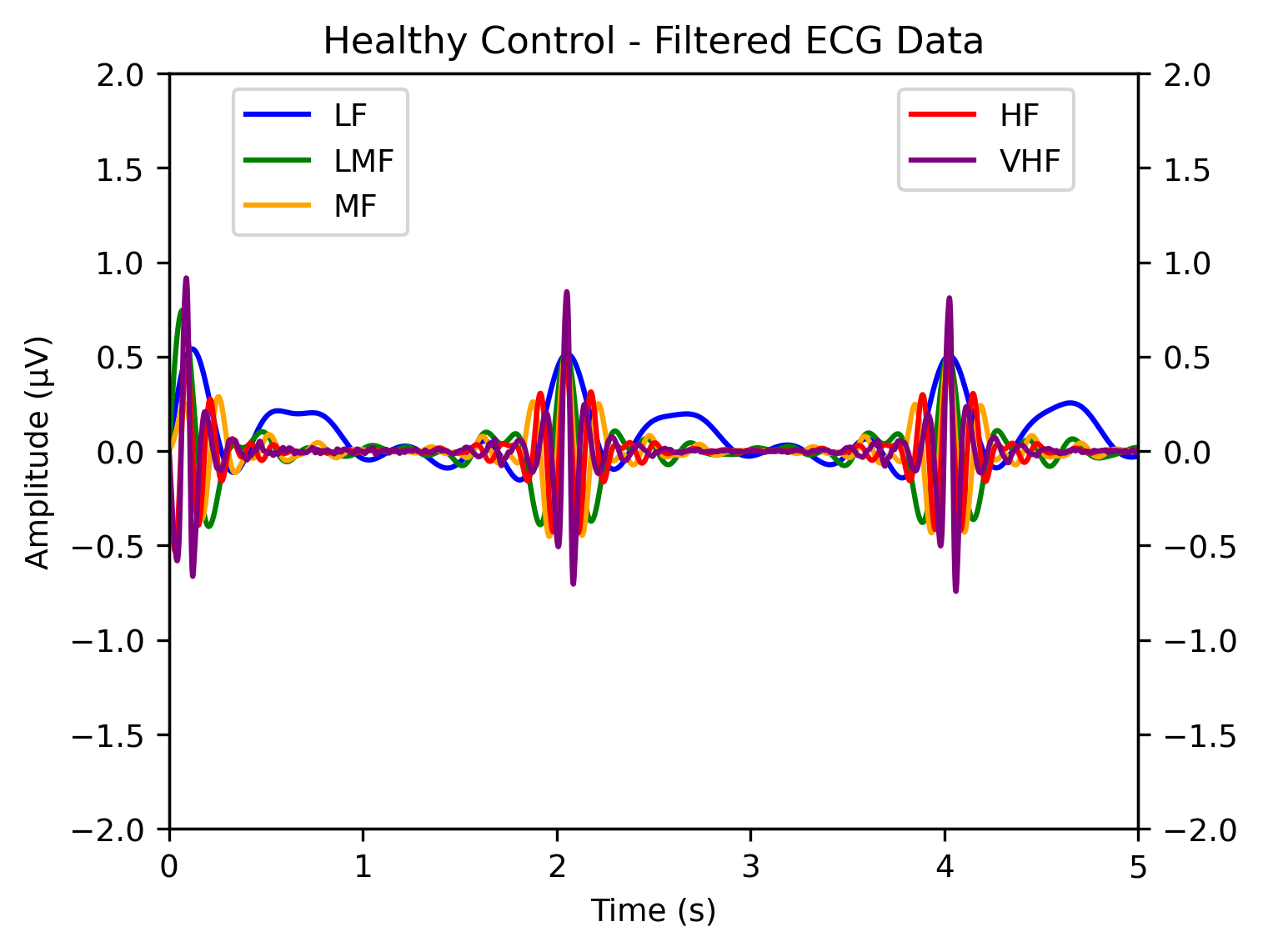} 
(D)\includegraphics[width = 3in]{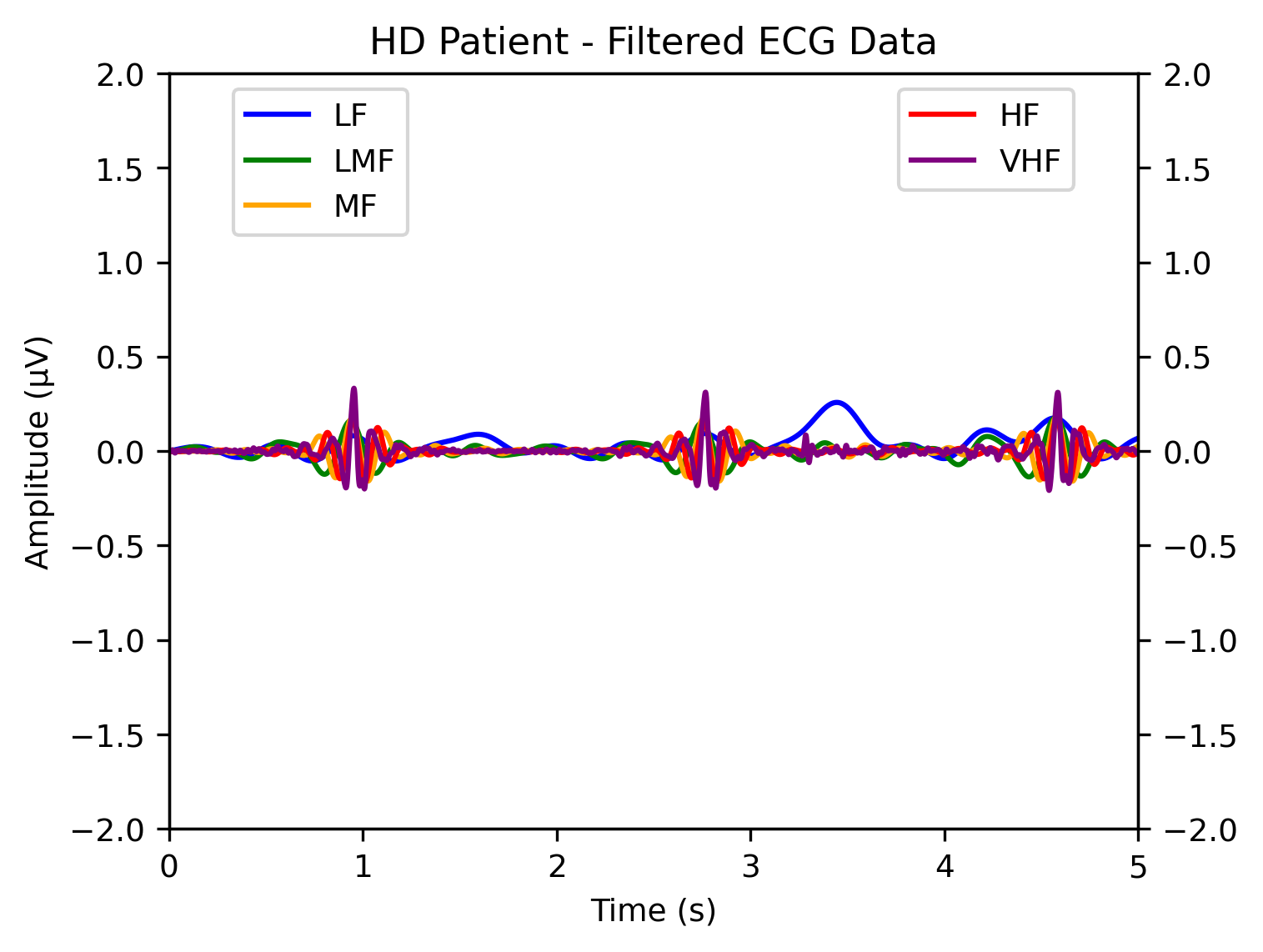}\\
(E)\includegraphics[width = 3in]{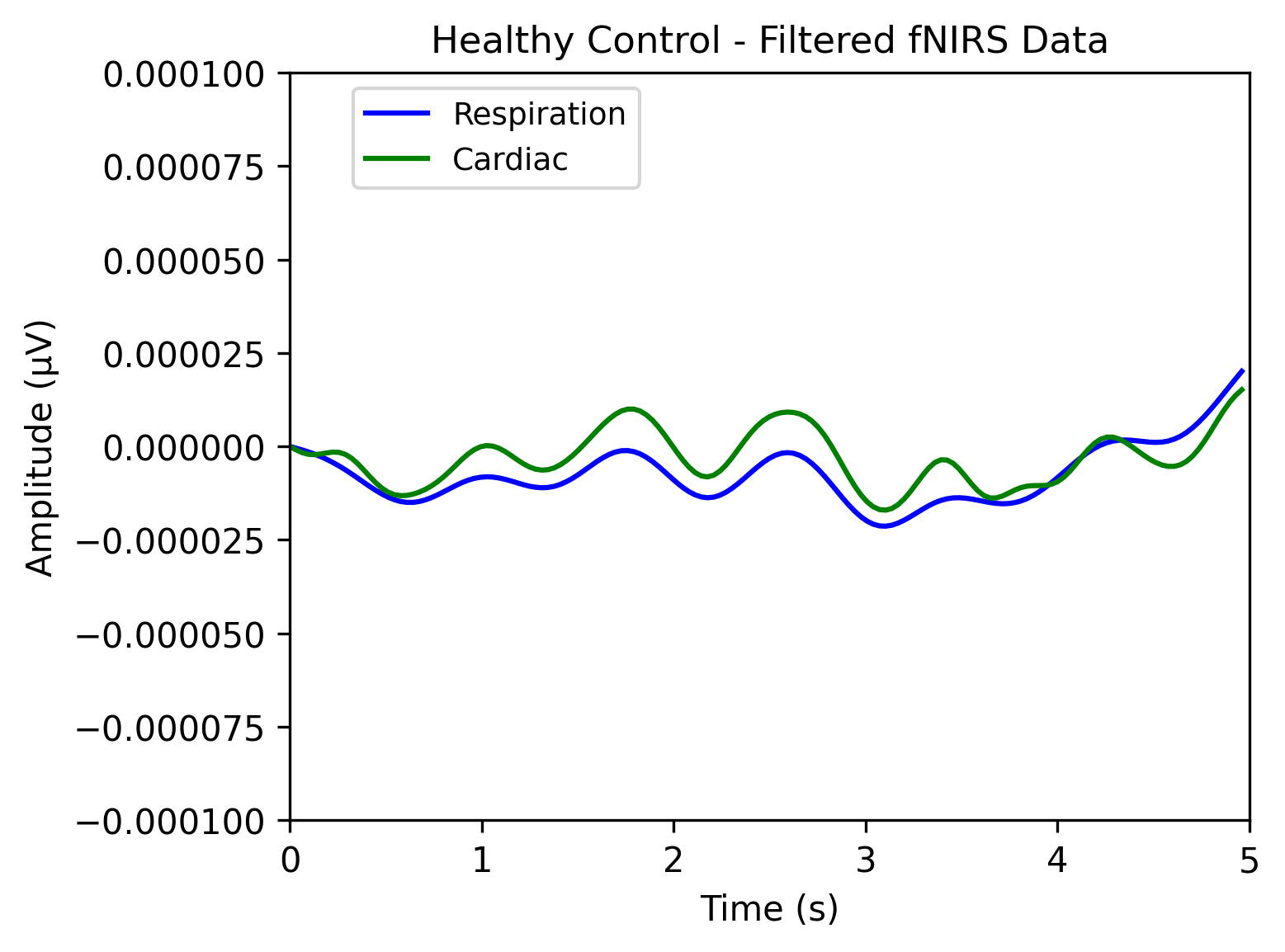}
(F)\includegraphics[width = 3in]{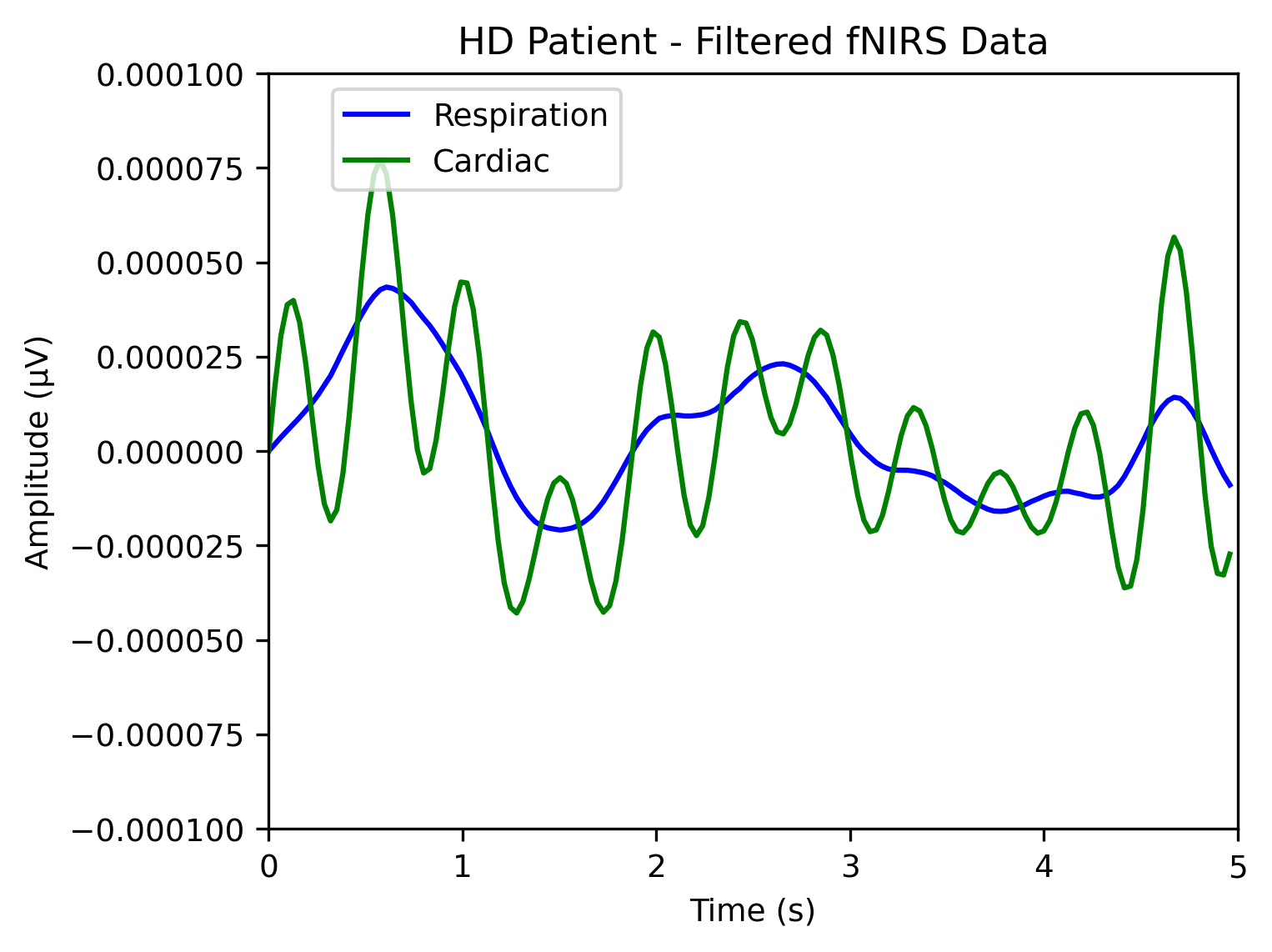}
\caption{Comparison of band-passed EEG, ECG, and fNIRS signals of healthy patients versus HD-positive patients. (A) Healthy EEG; (B) HD EEG; (C) Healthy ECG; (D) HD ECG; (E) Healthy fNIRS; (F) HD fNIRS.}
\label{fig:fig2}
\end{figure*}

\begin{figure*}
(A)\includegraphics[width = 6in]{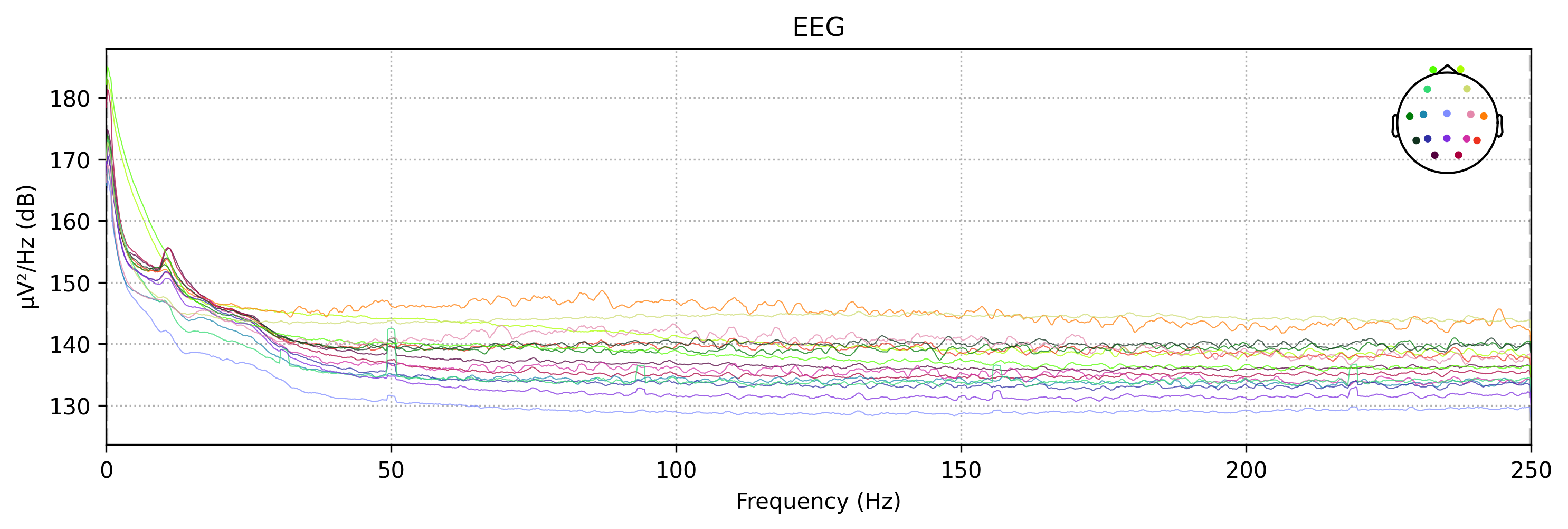}
(B)\includegraphics[width = 6in]{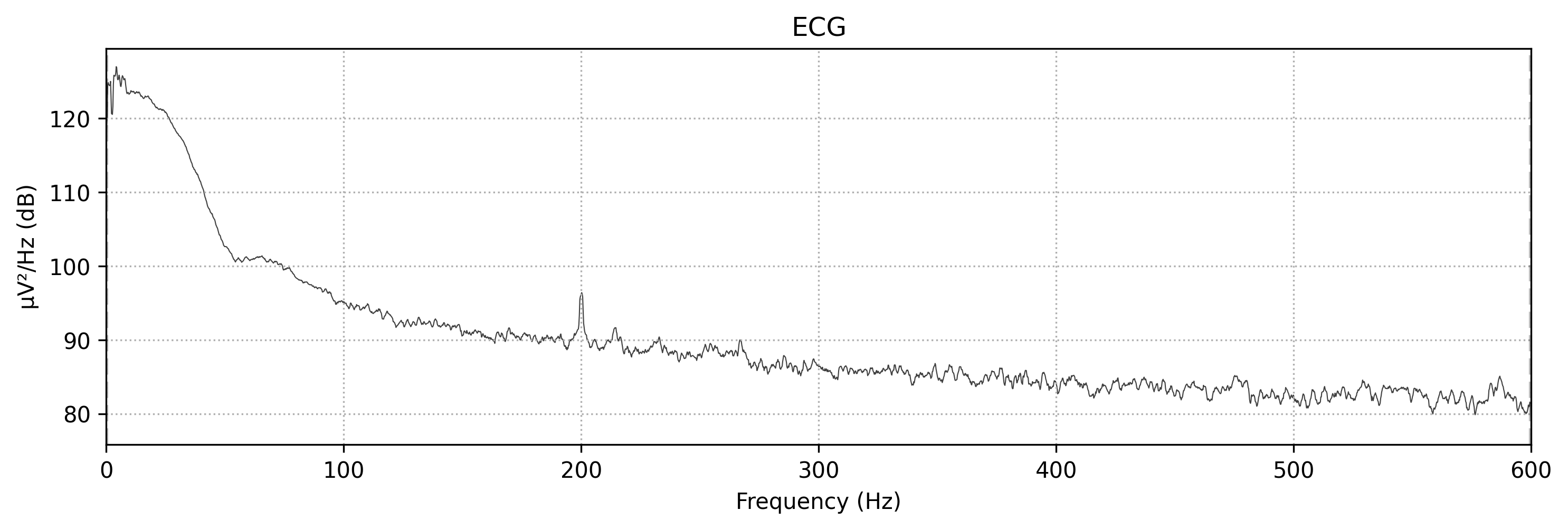}\\
(C)\includegraphics[width = 6in]{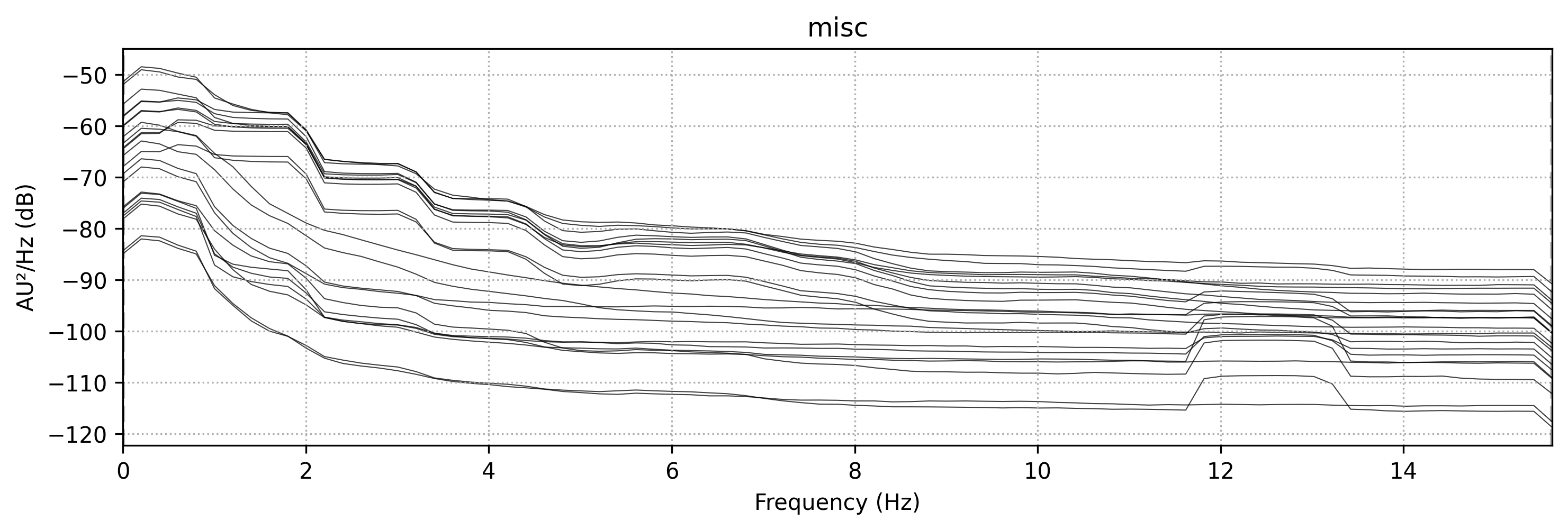}
\caption{EEG, ECG and fNIRS unfiltered power spectral density examples. (A) EEG PSD; (B) ECG PSD; (C) fNIRS PSD.}
\label{fig:fig3}
\end{figure*}

\hspace{0.5cm}Next, the data filtered and segmented. First, the data had to be converted into an MNE object, which was done in accordance with the 10-20 standard system (which the electrodes are placed in) for the EEG and fNIRS data. The EEG signal was filtered with EEG reference at a low pass of 0.5 Hz and a high pass of 45 Hz, in order to place an emphasis on the lower frequencies around the delta, theta, and alpha frequency bands, in which abnormalities can be indicative of HD pathology \cite{alphathetaborder}. The ECG data was filtered with low pass of 0.05 Hz (the lowest frequency recorded by the machine), and a high pass of 100 Hz to eliminate high-frequency noise \cite{ecgfilter}. The fNIRS signal was filtered at a low pass of 0.2 Hz and high pass of 1.5 Hz to remove artifacts caused by blood pressure fluctuations \cite{fnirsfilter}. The filtered signals for EEG, ECG, and fNIRS were then segmented into 5s epochs with 1s overlap, with the 1200s data for each patient being converted into 300 epochs--with 4s of unique data in each. For each patient, the bad epochs were dropped and the data was normalized by subtracting the mean of each channel. The data arrays for each signal type for all patients were saved in NPY format. An example of the EEG signal for each filtered band (same frequency bounds as shown for EEG in Table~\ref{tab:tab2}) for a healthy patient and HD patient are shown in Figure~\ref{fig:fig2}. As can be seen, the signal for the HD patient has generally lower amplitudes (in microvolts). In similar fashion, a filtered ECG signal example for healthy and HD patient can be seen in Figure~\ref{fig:fig2}, where the ECG bands are more consistent in amplitude and more periodic for the healthy control than in the HD patient. Figure~\ref{fig:fig2} shown a similar comparison for the filtered fNIRS signal, where both bands more closely follow each other for the healthy patient.
\hspace{0.5cm}Finally, the features could be extracted from the processed signals. As shown in Table~\ref{tab:tab2}, standard statistical and slope values were first extracted from the signals of each patient, namely kurtosis, coefficient of variation, skewness, 1st and 2nd difference of mean, 1st and 2nd difference of max, slope mean, slope variance, and Higuchi fractal dimension (HFD). They were extracted on a per-channel basis, thus amounting to 160 features from EEG, 10 features from ECG, and 220 features from fNIRS. HFD can give information about the complexity of neuronal activity in various neurophysiological conditions, and is simple to calculate \cite{hfd}. 
\begin{table*}[]
\centering
\caption{Extracted features. In the ECG power spectrum analysis, LF: Low frequency, LMF: Lower-middle frequency, MF: Middle frequency, HF: High frequency, VHF: Very-high frequency.}
\label{tab:tab2}
\resizebox{6in}{!}{%
\begin{tabular}{@{}lllllll@{}}
\toprule
\textbf{Hijorth} & \textbf{Statistical} & \textbf{Slope}       & \textbf{Wavelet} & \textbf{PSD (EEG)} & \textbf{PSD (ECG)} & \textbf{PSD (fNIRS)} \\ \midrule
Activity         & Kurtosis             & Slope Mean           & Approx. Mean     & \(\delta \) (0.5-3.5 Hz) & LF (0.05-6 Hz)     & Resp. (0.2-0.6 Hz)   \\
Mobility         & 2nd Diff. of Mean    & Slope Variance       & Approx. SD       & \(\theta \) (3.5-7.5 Hz) & LMF (6-11 Hz)      & Cardiac (0.6-1.5 Hz) \\
Complexity       & 2nd Diff. of Max     & Higuchi Fractal Dim. & Approx. Energy   & \(\alpha \) (7.5-13 Hz)  & MF (11-16 Hz)      &                      \\
                 & Skewness             &                      & Approx. Entropy  & \(\beta \) (13-30 Hz)    & HF (16-20 Hz)      &                      \\
                 & Coef. of Variation   &                      & Detailed Mean    & \(\gamma \) (30-45 Hz)   & VHF (20-100 Hz)     &                      \\
                 & 1st Diff. of Mean    &                      & Detailed SD      &                    &                    &                      \\
                 & 1st Diff. of Max     &                      & Detailed Energy  &                    &                    &                      \\
                 &                      &                      & Detailed Entropy &                    &                    &                      \\ \bottomrule
\end{tabular}%
}
\end{table*}
\hspace{0.5cm}Next, Hijorth parameters were extracted from each channel: activity, mobility, and complexity, totaling to 48 features per epoch for EEG, 3 features for ECG, and 66 features for fNIRS. Activity represents the variance or power of the signal over a certain epoch/segment in time domain, and indicative of the power spectrum surface over the frequency domain. Activity can be calculated for signals simply through:
\begin{equation}
    Activity=var(y(t))
\end{equation}
Mobility indicates the mean frequency over an epoch in the time domain, also represented as the proportion of standard deviation (SD) over the power spectrum, calculated through the following equation:
\begin{equation}
    Mobility=\sqrt{\frac{var(\frac{dy(t)}{dt})}{Activity(y(t))}}
\end{equation}
Lastly, complexity estimates the bandwidth of the signal over an epoch, essentially the average power of the second derivative of the signal, represented in:
\begin{equation}
    Complexity=\frac{Mobility(\frac{dy(t)}{dt})}{Mobility(y(t)))}
\end{equation}
\hspace{0.5cm}Several wavelet features were also extracted per channel, calculated using PyWavelets (https://pywavelets.readthedocs.io/) \cite{pywt}. To calculate each feature, a discrete wavelet transform (DWT) was performed and the coefficients from the wavelet function stored. For EEG and fNIRS, the Coiflet (coif1) wavelet was used, and for ECG, the Daubechies (db4) wavelet was used for features. Then, the approximate and detailed mean, SD, energy, and entropy were all calculated from these coefficients over a time segment, amounting to 128 EEG features, 8 ECG features, and 176 fNIRS features per epoch. Energy (4) is the summation of the squared signal, and entropy (5) is the measure of regularity/fluctuation in a time segment, shown in the following equations:
\begin{equation}
    Energy_{s}=\sum_{i=0}^{k}C_{s}^{2}
\end{equation}
\begin{equation}
    Entropy_{s}=\log C_{s}^{2}\cdot\sum_{i=0}^{k}C_{s}^{2}
\end{equation}
\hspace{0.5cm}The power-spectral density (PSD) values were calculated for a variety of bands based on frequency bounds shown in Table~\ref{tab:tab2}. The total amount of features amounted to 80 for EEG, 5 for ECG, and 44 for fNIRS. The ECG PSD bands were chosen based on the peak powers of typical ECG signals, and the fNIRS PSD bands were chosen based on respiration and heartbeat frequencies \cite{ecgbands,fnirsbands}. PSD indicates the power levels over the frequency domain in each component in a signal segment, which tells the model the range of power over various signal frequencies, making it an effective predictor for abnormalities \cite{EEGfeatures}. Welch's method of calculating PSD is an alternative approximation method to the Fast-Fourier Transform (FFT), which averages periodograms over time on a per-segment basis, represented through the equation \cite{welch}:
\begin{equation}
    PSD^{_{x}^{Welch}}(\omega _{k})_{=}^{\Delta }\frac{1}{K}\sum_{m=0}^{K-1}P_{x_{m}},M(\omega _{k})
\end{equation}
An example of PSD over the frequency domain for each signal type is shown in Figure~\ref{fig:fig3}. The final data array amounted to 20631 epochs from 69 patients with 948 features per epoch. A group array was also made to ensure that the epochs of each patient were trained or tested together, preventing overfitting.
\subsection{Model Selection and Construction}
\hspace{0.5cm}Classical machine learning algorithms followed by deep neural networks were used for classification tasks based on the extracted EEG features. All classical machine learning algorithms were implemented using Scikit-learn (https://scikit-learn.org/), including Random Forest, Extremely Randomized Trees, Logistic Regression, Support-vector machines, Linear Discriminant Analysis (LDA), and Quadratic Discriminant Analysis (QDA) \cite{scikit-learn}. All models were tested using 10-fold cross validation with the group array to minimize overfitting and underfitting, and scored with accuracy, precision, recall, receiver operating characteristic (ROC-AUC), and F1 score. The best models were then found and hyperparameter-tuned through random grid search.

\hspace{0.5cm}Random Forest is an ensemble method that uses many randomized decision trees, each of which are trained/fit on sub-samples of the dataset and subsets of the features at each split point, with their results averaged. This robust approach of different features at each tree split point has been shown to improve accuracy, reduce variability and reduce overfitting. Random Forests have been popular in qEEG analysis, such as in differentiating schizophrenia patients from controls \cite{Schizo}. 

\hspace{0.5cm}Extremely Randomized trees (ERTs) are an ensemble method in of themselves, and is similar to Random Forest in that ERTs take subsets of the features to train each sub-tree, however contrast in that ERTs train each sub-tree on the entire dataset instead of a sub-sample. ERTs can also select either a random split or the best split for the optimal split, unlike the greedy algorithm of Random Forests. 

\hspace{0.5cm}Logistic Regression is effectively a linear regressor that models the data using a sigmoid function instead of a linear function. The algorithm itself remains similar to a linear regressor until the decision threshold, where in binomial logistic regression, the probabilistic result is reduced into a binary output (0 or 1).

\hspace{0.5cm}Support-Vector machines (SVM) use hyper planes to effectively partition data into classes, and are thus better suited for classification rather than regression. Based on the number of features, the algorithm finds, in an N-dimensional plane, an equation of N-1 dimensions that partitions the data into various classes, in this case two classes (0 or 1), through minimizing the distance between the equation graph and the individual data points (functional margin). Preliminary tests showed that the polynomial kernel outperformed other kernels, such as sigmoid and radial-basis-function, and thus was used for the remainder of the study. 

\hspace{0.5cm}Linear Discriminant Analysis uses a similar method to SVM, attempting to draw planes to divide sets of classes in the hyperplane of features, however uses different criteria. LDA attempts to maximize the mean distance between the points and the hyperplane, and minimize the variance between each class. Thus LDA performs similarly to SVM. QDA uses the same criteria, but with a degree-2 discriminant plane rather than linear.

\section{Results}
\label{sec:results}
\subsection{Preliminary Model Results}
\hspace{0.5cm}The models outlined were first run with default hyperparameters testing a variety of metrics with k=10 on a personal computer with Intel Core i7-3930k CPU and NVIDIA GeForce RTX-2060 GPU. The concatenated feature vectors were created along with the label and group arrays, and fed into cross validation. The results of the models are shown in Table~\ref{tab:tab3}. As can be seen, Extremely Randomized Trees (ERTs) performed best with Random Forest close behind on all metrics except recall and f1 score, where the Random Forest outperformed and tied the ERTs, respectively. In the next section, the estimators of the Random Forest and ERT models were scaled out.
\begin{table*}[]
\centering
\caption{Results for binary classification with all models using default hyperparameters. Tested with k=10 with group array with all epochs of all patients (n=20631).}
\label{tab:tab3}
\resizebox{6in}{!}{%
\begin{tabular}{@{}llllll@{}}
\toprule
\textbf{Model} & \textbf{Accuracy (\%)} & \textbf{Precision (\%)} & \textbf{Recall (\%)} & \textbf{ROC AUC} & \textbf{F1 score} \\ \midrule
ERT            & 90.642                & 87.374                 & 85.994              & 0.937              & 0.853               \\
Random Forest  & 89.400                & 86.209                 & 86.971              & 0.936              & 0.853               \\
SVM            & 62.649                & 55.977                 & 2.611              & 0.564              & 0.047               \\
Log. Regr.     & 66.124                & 59.456                 & 42.599              & 0.615              & 0.467               \\
LDA            & 81.027                & 73.889                 & 84.431              & 0.875              & 0.765               \\
QDA            & 67.971                & 57.286                 & 79.620              & 0.747              & 0.658               \\ \bottomrule
\end{tabular}
}
\end{table*}
\subsection{Optimized Results}
\begin{figure*}
(A)\includegraphics[width = 3in]{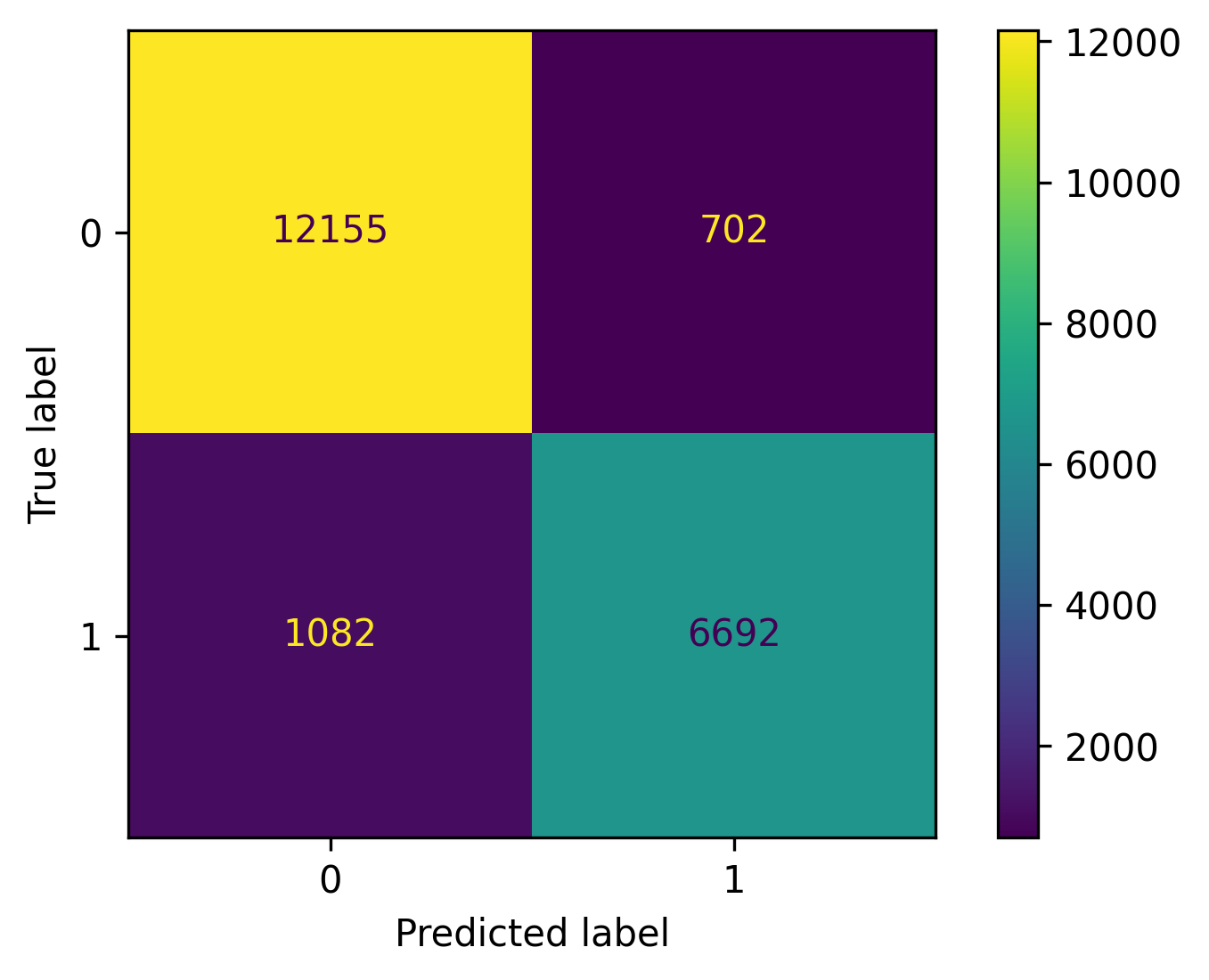}
(B)\includegraphics[width = 3in]{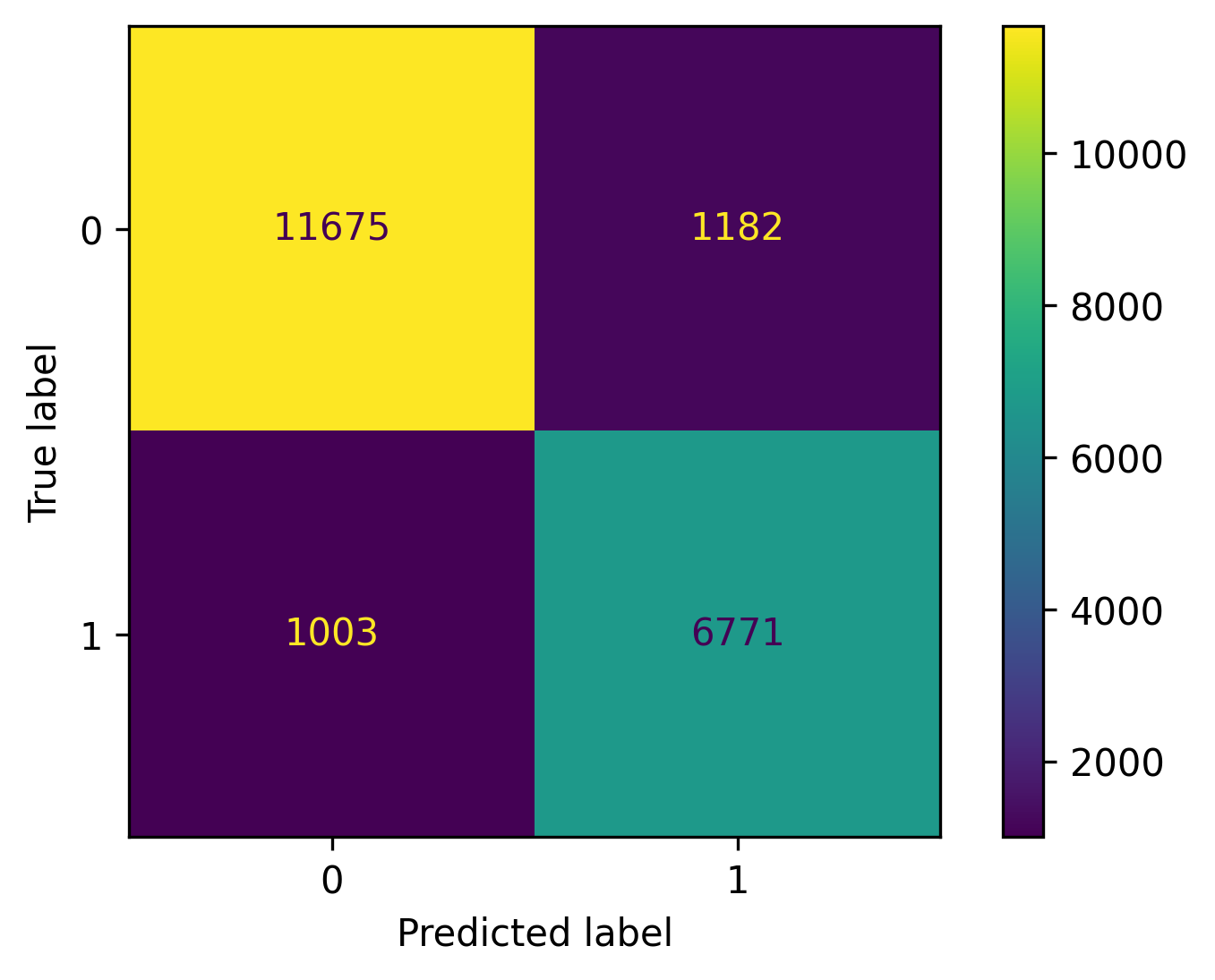}
\caption{Confusion matrices for tuned models. (A) Tuned ERT; (B) Tuned Random Forest.}
\label{fig:fig4}
\end{figure*}

\begin{figure*}
(A)\includegraphics[width = 3in]{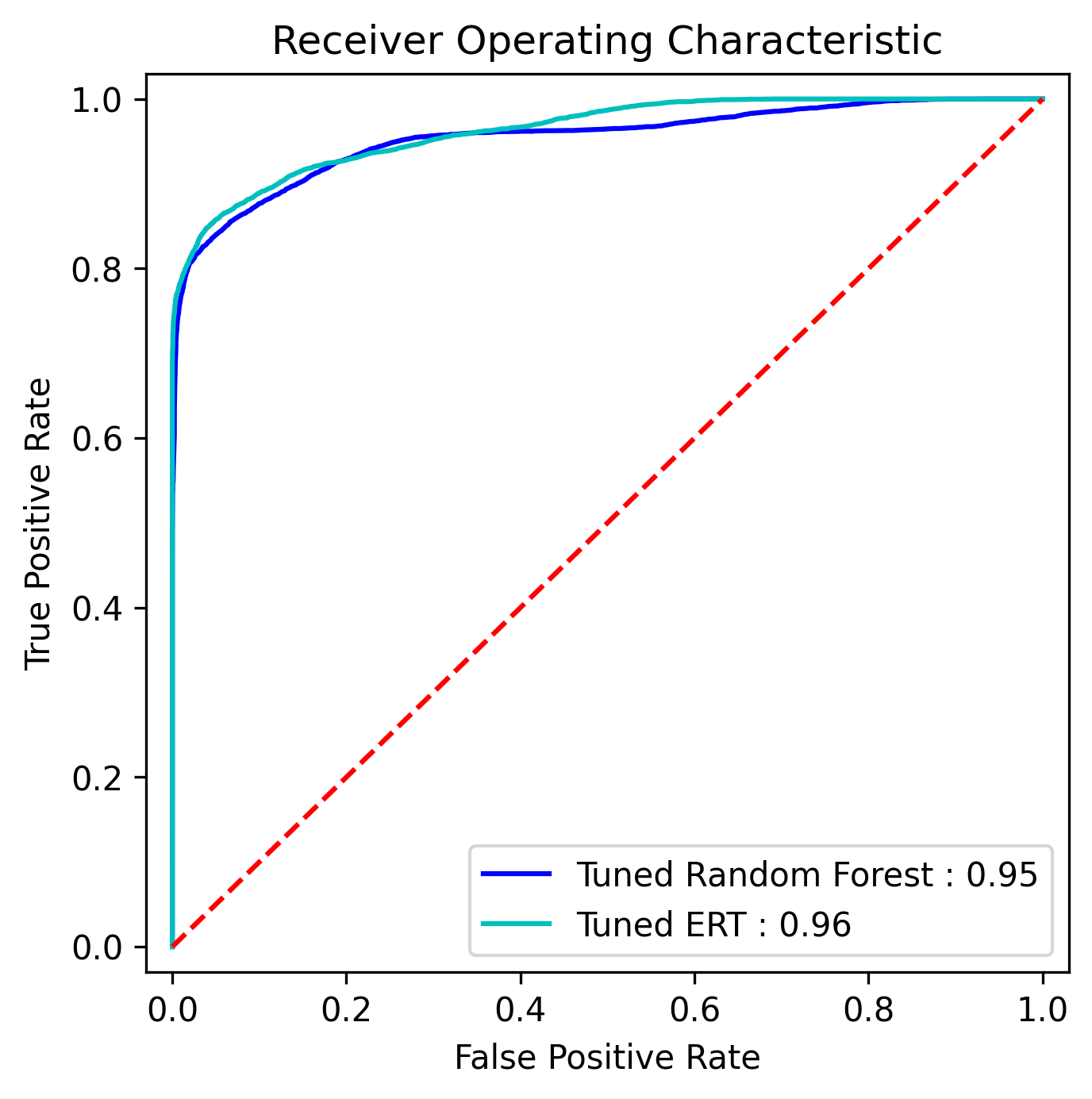}
(B)\includegraphics[width = 3in]{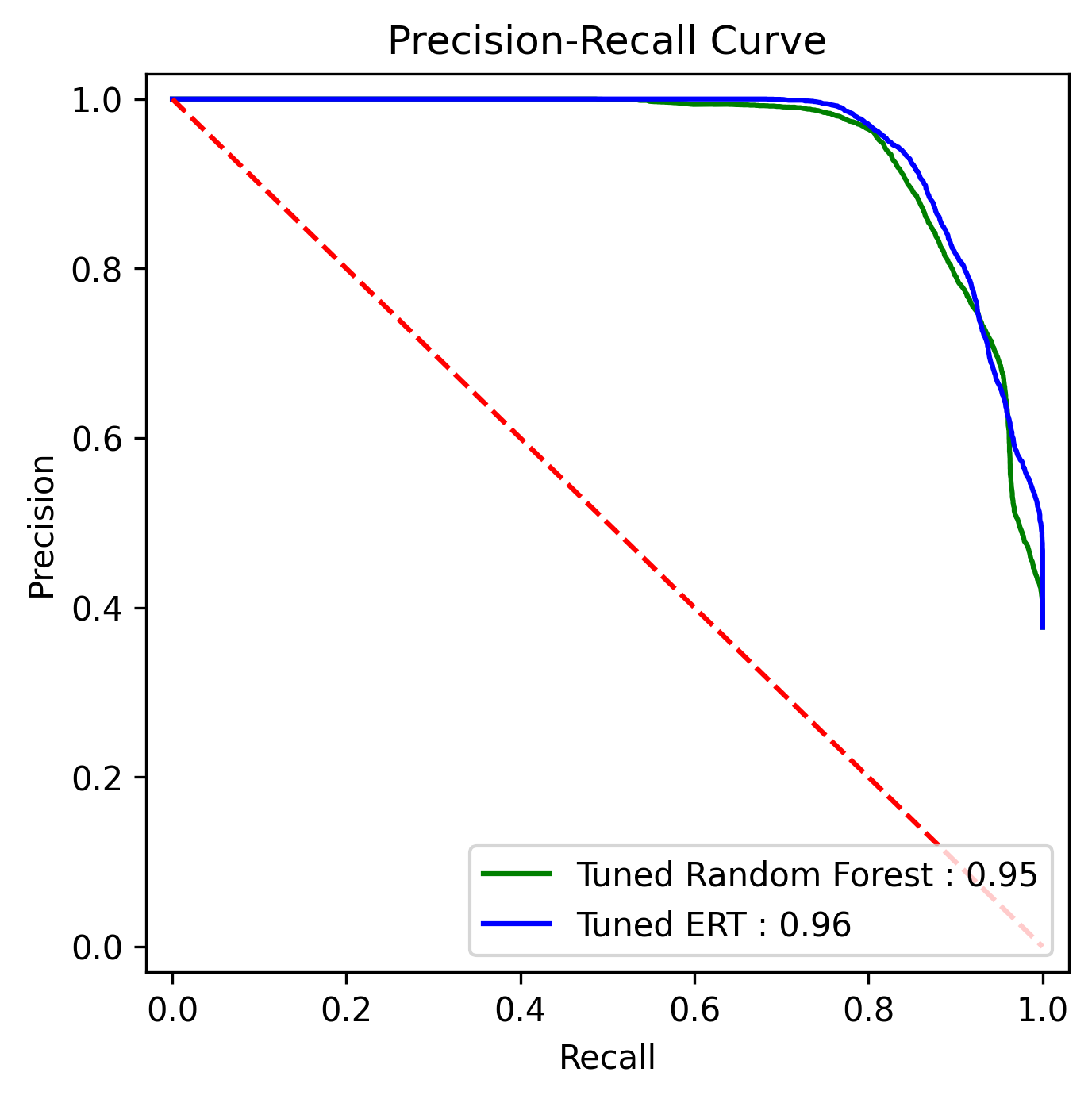}
\caption{ROC area under the curve (AUC) and precision-recall AUC for tuned models. (A) ROC; (B) Precision-Recall.}
\label{fig:fig5}
\end{figure*}

\hspace{0.5cm}Here, the ERT and Random Forest algorithms were scaled out to better accommodate for the large feature count and sample size. The estimator counts were increased to n=1000 estimators in an experiment to evaluate the performance benefit of large estimator counts with many features. The two tuned models were them run on the same cross validation evaluator with k=10 and the full dataset, the results for which are shown in Table~\ref{tab:tab4}. Once again, the Extremely Randomized Trees algorithm performed the best, again on all metrics except for recall.
\begin{table*}[]
\centering
\caption{Results for binary classification with tuned ERT and Random Forest models using estimators=1000. Tested with k=10 with group array with all epochs of all patients (n=20631).}
\label{tab:tab4}
\resizebox{6in}{!}{%
\begin{tabular}{@{}llllll@{}}
\toprule
\textbf{Model}      & \textbf{Accuracy (\%)} & \textbf{Precision (\%)} & \textbf{Recall (\%)} & \textbf{ROC AUC} & \textbf{F1 score} \\ \midrule
Tuned ERT           & 91.353                & 90.506                 & 86.082              & 0.963              & 0.882               \\
Tuned Random Forest & 89.409                & 85.138                 & 87.098              & 0.953              & 0.861               \\ \bottomrule
\end{tabular}%
}
\end{table*}
\hspace{0.5cm}The results for the tuned models were visualized in a variety of ways. First confusion matrices showing the false-positives, true-positives, false-negatives, and true-negatives were created from the prediction arrays of the ERT and Random Forest evaluation, shown in Figure~\ref{fig:fig4}. Next, area-under-the-curve (AUC) plots were created from the false-positives and true-positive rates to visualize the receiver operating characteristic, as well as from the precision and recall for the precision-recall curve, shown in Figure~\ref{fig:fig5}.
\subsection{Statistical analysis}
\begin{figure*}
(A)\includegraphics[width = 3in]{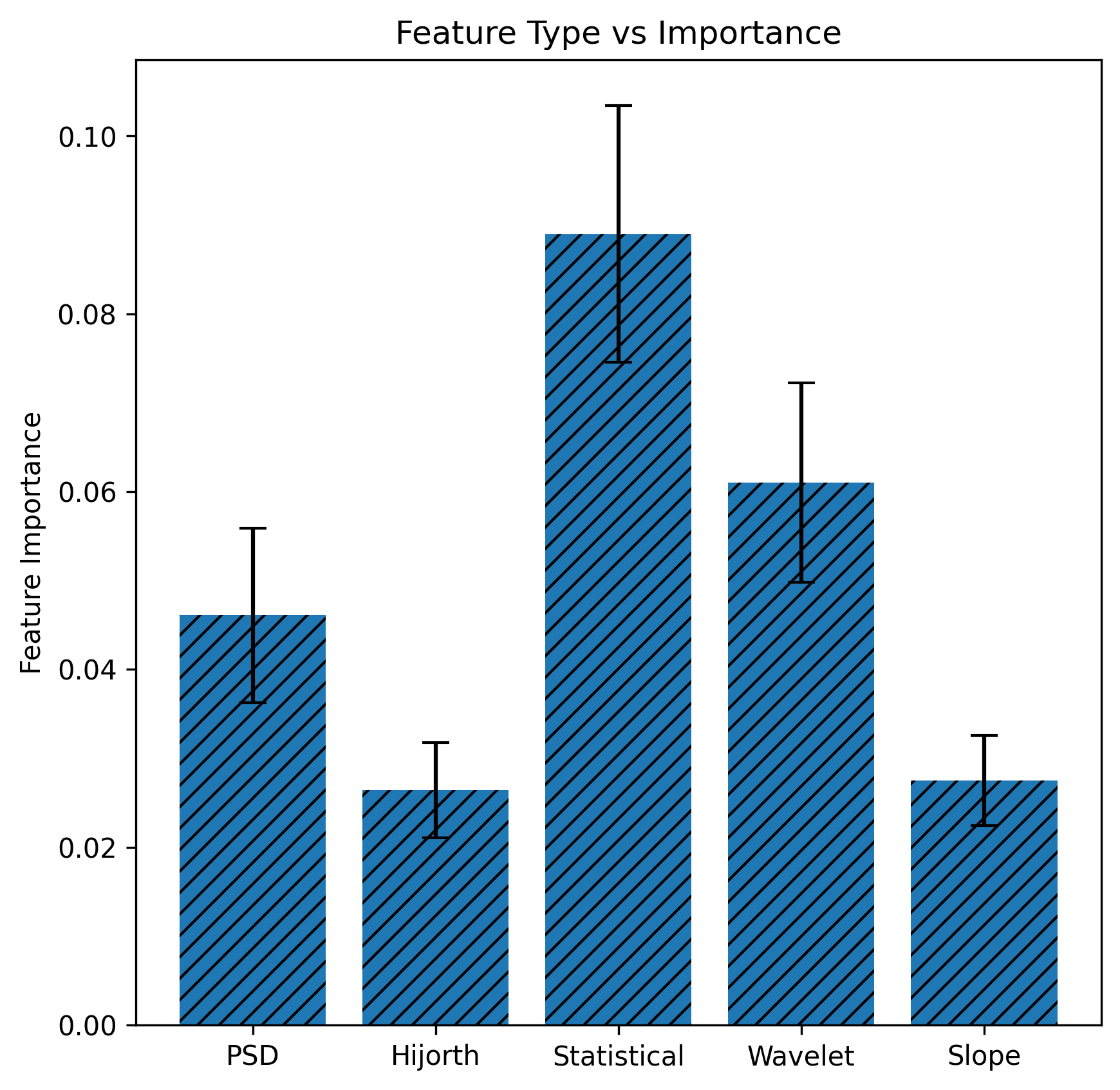}
(B)\includegraphics[width = 3in]{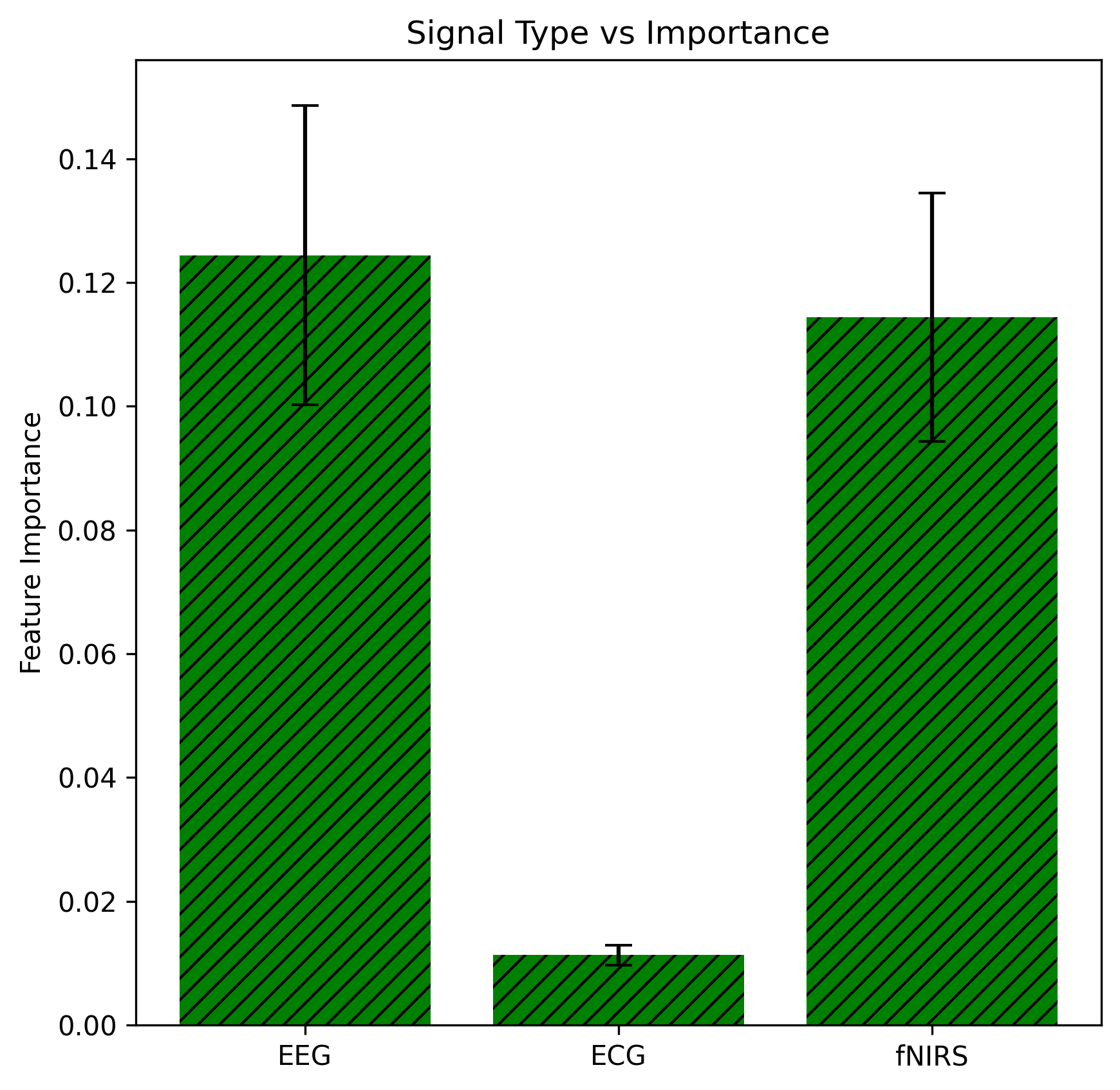}

(C)\includegraphics[width = 3in]{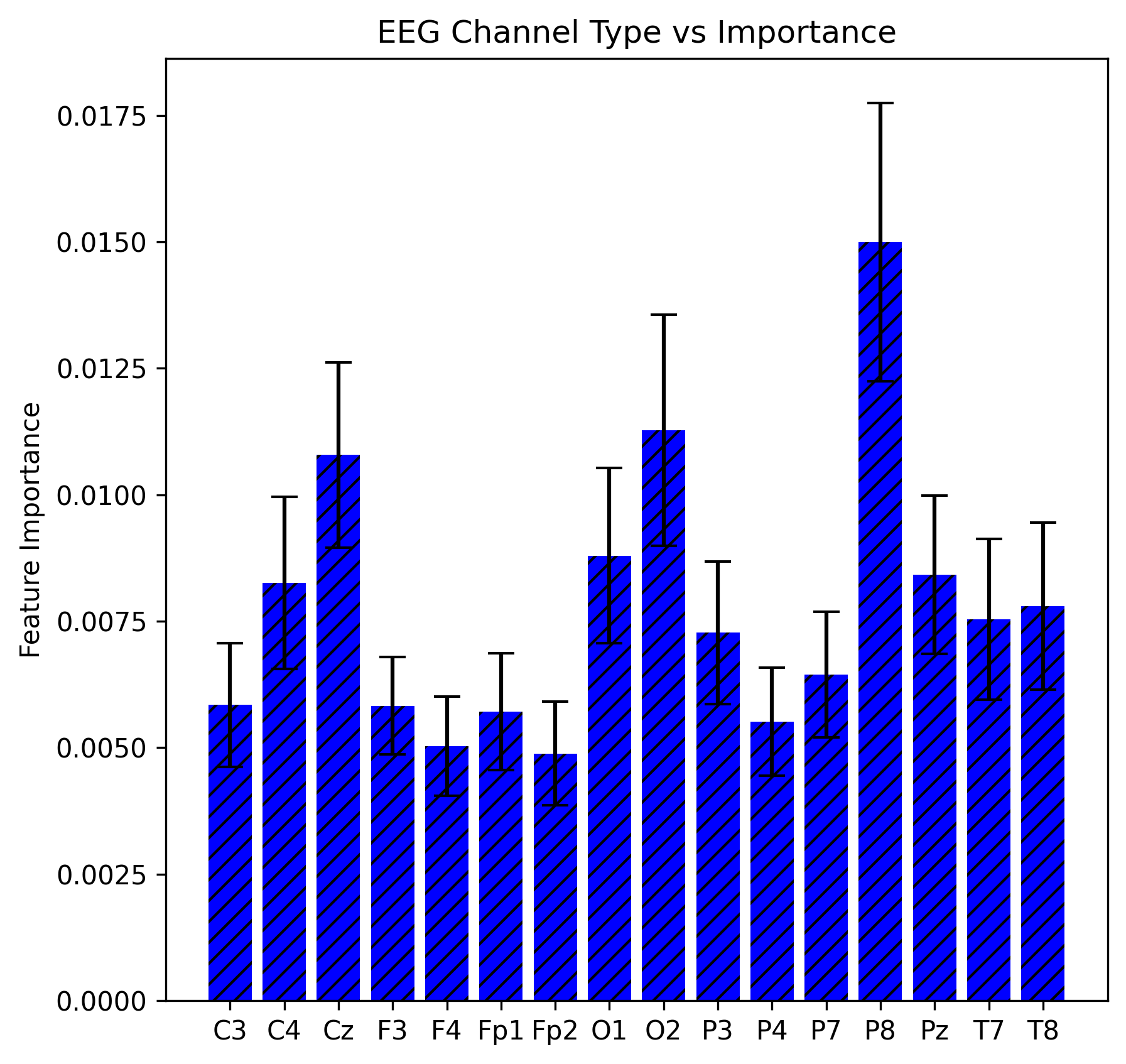}
(D)\includegraphics[width = 3in]{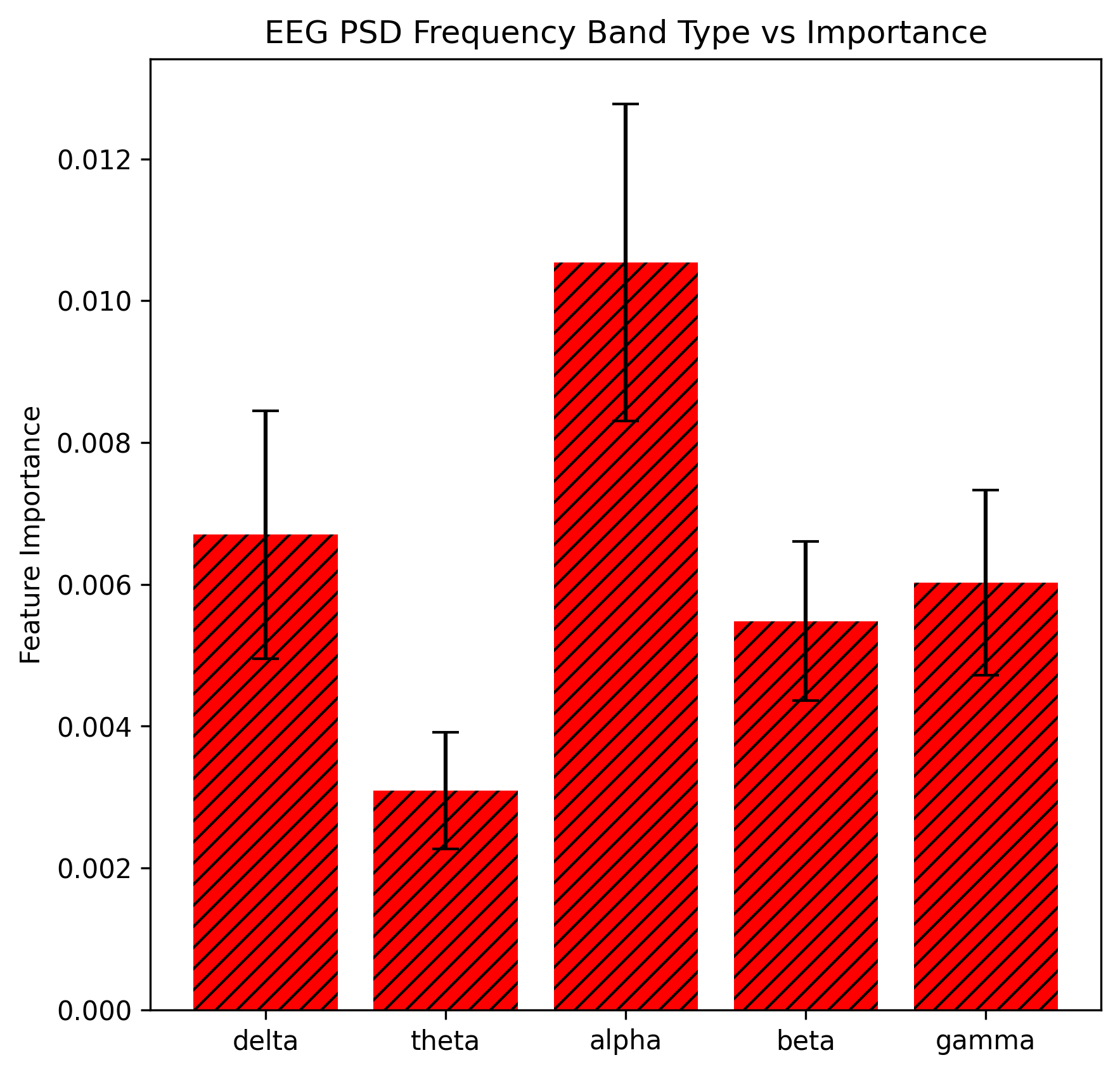}
\caption{Comparison of feature importance $\pm$ 1 SE based on various subsets of features (A) Basis of feature; (B) Basis of signal; (C) Basis of EEG channel; (D) Basis of EEG PSD band.}
\label{fig:fig6}
\end{figure*}

\hspace{0.5cm}Statistical significance was also calculated via a variety of tests for the features using Statsmodels (https://www.statsmodels.org/) \cite{statsmodels}. For the entire feature array, the R\textsuperscript{2} coefficient of determination---representing the proportion of variation of the dependent variable explainable by the independent variable---was calculated to be 0.89. The F-statistic was calculated to be 275.9 and the probability-F-statistic was 0, representing the probability that the observed results occurred by chance, ie. approximately 0\%. The log-likelihood, or an estimate for the goodness-of-fit for the data and labels, came out to be 8412. More specific values, such as p and t values, were calculated on a per-feature basis with the results summarized in Table~\ref{tab:tab5}.
\begin{table*}[]
\centering
\caption{Variety of statistical-significance values for various p value thresholds on a feature-by-feature basis (n=948).}
\label{tab:tab5}
\resizebox{6in}{!}{%
\begin{tabular}{@{}lllll@{}}
\toprule
\textbf{p value range} & \textbf{Count (\#)} & \textbf{p value} & \textbf{t value} & \textbf{Std. error} \\ \midrule
p = 0                    & 357                 & 0 $\pm$ 0             & -0.825 $\pm$ 8.049    & 240.217 $\pm$ 586.147   \\
p \textless{} 0.001      & 22                  & 0.001 $\pm$ 0         & -0.878 $\pm$ 3.277    & 252.599 $\pm$ 632.387    \\
p \textless{} 0.01       & 76                  & 0.00499 $\pm$ 0.0024  & -0.130 $\pm$ 2.853    & 324.256 $\pm$ 690.298   \\
p \textless{} 0.05       & 122                 & 0.0277 $\pm$ 0.0117   & -0.263 $\pm$ 2.224    & 168 $\pm$ 462            \\
p \textgreater{} 0.05    & 371                 & 0.421 $\pm$ 0.275     & 0.0669 $\pm$ 1.068    & 167.394 $\pm$ 465.572   \\ \bottomrule
\end{tabular}%
}
\end{table*}
\hspace{0.5cm}Feature importance was also calculated in order to reveal information into the most important values for predictions. Since performance was not particularly important, the feature importance was calculated through a default-hyperparameterized ERT. The data was split up on the basis of various attributes, such as the general group of feature, the type of signal, the specific channels, and bands. The results are shown for each in Figure~\ref{fig:fig6}.
\section{Discussion}
\label{sec:discussion}

\hspace{0.5cm}Overall, the procedure seem to have been chosen appropriately given the observations. For example, the EEG signal visualization in Figure~\ref{fig:fig3}.a, all of the power of the signal seems to be contained in Hz$<$50, with severe fluctuations at Hz=50. In the ECG signal, almost all the power is contained within Hz$<$~60, and most of the power of fNIRS is contained before ~4 Hz---however most of the fNIRS activity occurs within Hz$<$2.0. These observations did not change substantially with different patient examples, indicating that the filtering bounds chosen for the three signal types were appropriate. Additionally, the choice of k-folds of 10 for 10-fold cross validation seems to have resulted in a model that is neither extremely underfitted nor extremely overfitted, as is usually the case with such a value of k. 

\hspace{0.5cm}In this study, the relationship between EEG, ECG, and fNIRS and the state of a patient as HD-afflicted or healthy was statistically significant and consistent. In terms of statistical significance, as shown in Table~\ref{tab:tab5}, 577 out of the 948 total features had p$<$0.05, signifying statistical significance, with 357 of them having negligible p values. For these features, the null hypothesis can be rejected, and the null hypothesis can also safely be rejected when questioning the significance of the feature choice as a whole. T values---the difference of mean between groups, used to assess covariance---was high for the most statistically significant features. 

\hspace{0.5cm}The feature importance observations reveal some insight into the values most indicative of HD. Specifically, while the EEG signal was the most significant signal type by a small margin, the error bars have too large an overlap to conclude the feature importance of fNIRS in comparison to EEG---however ECG can be concluded to be the least significant signal type. In terms of feature type, the statistical and wavelet features were the most significant, which could call for more experimentation into various wavelet transform functions. Another interesting observation is that the P8 channel type was the most significant while the P7 channel had one of the lowest importances, despite both electrodes being symmetrical on either side of the head---which could indicate an asymmetry of HD. The feature importances in the EEG PSD frequency bands were mostly as expected, except for the apparent unimportance of the theta band, which could indicate some amount of noise in the theta band frequencies.

\hspace{0.5cm}The R\textsuperscript{2} coefficient of determination showed that 89\% of the variation of the dependent variable (HD diagnosis) was predictable by the independent variable(s), which follows closely with the accuracy of the two best models, Extremely Randomized Trees and Random Forest. The probability f-statistic also shows a negligible probability of lucky accuracy. Overall, the feature choice reflects a large amount of information about the patient's pathology with HD.

\hspace{0.5cm}Holistically, the models demonstrate the potential for biological signals such as EEG, ECG, and fNIRS for marking the path of Huntington's disease. The consistent superiority of the Extremely Randomized Trees algorithm was surprising considering the lack of its use in previous literature, such as in Schizophrenia \cite{discussionone}. The LDA also performed better than expected---achieving a high ROC AUC of 0.875---in comparison to its more complex QDA, suggesting more simple clustering divisions achievable by linear separators. The results of this study provide for a significant performance improvement over Odish et al. (2018), which used power spectral density of EEG data to classify Huntington's disease, achieving 83\% accuracy and 0.9 ROC AUC \cite{discussiontwo}.

\hspace{0.5cm}There are several limitations to this study. First, despite the analysis into feature importance, it is still difficult to predict which features will scale well into larger or more comprehensive datasets with more patients and electrode channels. Second, the 6 unclassified patients still prompt some investigation, although removing them from the dataset altogether did not notably affect performance. Third, this study did not evaluate the various markers tested at differing stages of the disease, which would have likely been useful in assessing the efficacy of HD patient monitoring through biological signals. These are all topics for future investigation.

\section{Conclusion}
\label{sec:conclusion}

\hspace{0.5cm}In this study we consolidated, preprocessed, and extracted features from a variety of biological signals from a complete dataset. We then formatted, filtered, segmented, and extracted features from the data, pertaining to hijorth, statistical, slope, wavelet, and power spectral features which had mostly statistically-significant correlations to HD positivity. We then ran a variety of machine learning algorithms and found high accuracy and ROC AUC with tree ensemble algorithms. We found the most significant features to be from EEG and fNIRS data, with statistical and wavelet features, with ECG being less useful. While there is still much research to be done in this area, this study provides some insight into the usefulness of neural and cardiac signaling for Huntington's disease prognosis, and could potentially be built upon with deeper or more powerful models, larger or more expansive datasets, or more thorough feature selection. Topics for future research could include evaluating biological signal features at various stages pre and post-Huntington's disease onset, as well as exploration of other neural signaling types such as MEG.

\section*{Acknowledgment}

We would like to thank Christine Yoon (B.S.) for her mentorship and support throughout this project.

\bibliographystyle{unsrt}
\bibliography{references}

\begin{thebibliography}{10}

\bibitem{HuntingtonDisease}
Anitha Ajitkumar and Orlando De~Jesus.
\newblock {\em Huntington Disease}.
\newblock 2022.

\bibitem{HuntingtonPrevalance}
Michael Rawlins, Nancy Wexler, Alice Wexler, Sarah Tabrizi, Ian Douglas,
  Stephen Evans, and Liam Smeeth.
\newblock The prevalence of huntington's disease.
\newblock {\em Neuroepidemiology}, 2016.

\bibitem{HuntingtonClinicals}
Peter McColgan and Sarah Tabrizi.
\newblock Huntington's disease: a clinical review.
\newblock {\em European Journal of Neurology}, 2018.

\bibitem{EarlyDetection}
Ipek Oguz.
\newblock Early detection of huntington’s disease: longitudinal analysis of
  basal ganglia and cortical thickness.
\newblock {\em National Institutes of Health; Project Number: 5R01NS094456-05},
  2011.

\bibitem{Currentdiagnosis}
Thomas Stoker, Sarah Mason, Julia Greenland, Simon Holden, Helen Santini, and
  Roger Barker.
\newblock Huntington's disease: diagnosis and management.
\newblock {\em Practical Neurology}, 2021.

\bibitem{BackgroundOne}
J.~Paulsen, D.~Langbehn, J.~Stout, E.~Aylward, C.~Ross, M.~Nance, M.~Guttman,
  S.~Johnson, M.~MacDonald, L.~Beglinger, K.~Duff, E.~Kayson, K.~Biglan,
  I.~Shoulson, D.~Oakes, and M.~Hayden.
\newblock Detection of huntington’s disease decades before diagnosis: the
  predict-hd study.
\newblock {\em Journal of Neurology, Neurosurgery, and Psychiatry}, 2008.

\bibitem{BackgroundThree}
Jane Paulsen.
\newblock Functional imaging in huntington’s disease.
\newblock {\em Experimental Neurology}, 2009.

\bibitem{BackgroundTwo}
Sarah Mason, Richard Daws, Eyal Soreq, Eileanoir Johnson, Rachael Scahill,
  Sarah Tabrizi, Roger Barker, and Adam Hampshire.
\newblock Predicting clinical diagnosis in huntington's disease: An imaging
  polymarker.
\newblock {\em Annals of Neurology}, 2018.

\bibitem{EEGThree}
D.~Scott, K.~Heathfield, B.~Toone, and J.~Margerison.
\newblock The eeg in huntington's chorea: a clinical and neuropathological
  study.
\newblock {\em Journal of Neurology, Neurosurgery, and Psychiatry}, 1972.

\bibitem{EEGFour}
Ksenija Cankar, Ziva Melik, Jan Kobal, and Vito Starc.
\newblock Evidence of cardiac electrical remodeling in patients with huntington
  disease.
\newblock {\em Brain and Behavior}, 2018.

\bibitem{EEGOne}
Nilesh Kulkarni and V.~Bairagi.
\newblock Diagnosis of alzheimer disease using eeg signals.
\newblock {\em International Journal of Engineering Research and Technology},
  2014.

\bibitem{EEGTwo}
Ana Paula, Maira Araujo, Maria Karoline, Juliana Carneiro, Marcelo Rodrigues,
  and Wellington Santos.
\newblock Early diagnosis of parkinson’s disease using eeg, machine learning
  and partial directed coherence.
\newblock {\em Research on Biomedical Engineering}, 2020.

\bibitem{data}
Juliane Bjerkan, Aneta Stefanovska, Gemma Lancaster, Jan Kobal, and Bernard
  Maglic.
\newblock Huntington's disease and controls dataset, 2021.

\bibitem{MNE}
Alexandre Gramfort, Martin Luessi, Eric Larson, Denis~A. Engemann, Daniel
  Strohmeier, Christian Brodbeck, Roman Goj, Mainak Jas, Teon Brooks, Lauri
  Parkkonen, and Matti~S. H{\"a}m{\"a}l{\"a}inen.
\newblock {{MEG}} and {{EEG}} data analysis with {{MNE}}-{{Python}}.
\newblock {\em Frontiers in Neuroscience}, 7(267):1--13, 2013.

\bibitem{alphathetaborder}
Natalya Ponomareva, Sergey Klyushnikov, Natalya Abramyacheva, Daria Malina,
  Nadejda Scheglova, Vitaly Fokin, Irina Ivanova-Smolenskaia, and Sergey
  Illarioshkin.
\newblock Alpha-theta border eeg abnormalities in preclinical huntington's
  disease.
\newblock {\em Journal of the Neurological Sciences}, 2014.

\bibitem{ecgfilter}
Rahul Kher.
\newblock Signal processing techniques for removing noise from ecg signals.
\newblock 2019.

\bibitem{fnirsfilter}
Franziska Klein and Cornelia Kranczioch.
\newblock Signal processing in fnirs: A case for the removal of systemic
  activity for single trial data.
\newblock {\em Frontiers in Human Neuroscience}, 13, 2019.

\bibitem{hfd}
Srdjan Kesić and Sladjana~Z. Spasić.
\newblock Application of higuchi's fractal dimension from basic to clinical
  neurophysiology: A review.
\newblock {\em Computer Methods and Programs in Biomedicine}, 133:55--70, 2016.

\bibitem{pywt}
Gregory~R. Lee, Ralf Gommers, Filip Waselewski, Kai Wohlfahrt, and Aaron Leary.
\newblock Pywavelets: A python package for wavelet analysis.
\newblock {\em Journal of Open Source Software}, 4(36):1237, 2019.

\bibitem{ecgbands}
J.~McNames, C.~Crespo, M.~Aboy, J.~Bassale, L.~Jenkins, and B.~Goldstein.
\newblock Harmonic spectrogram for the analysis of semi-periodic physiologic
  signals.
\newblock In {\em Proceedings of the Second Joint 24th Annual Conference and
  the Annual Fall Meeting of the Biomedical Engineering Society] [Engineering
  in Medicine and Biology}, volume~1, pages 143--144 vol.1, 2002.

\bibitem{fnirsbands}
Md.~Asadur Rahman, Mohd~Abdur Rashid, and Mohiuddin Ahmad.
\newblock Selecting the optimal conditions of savitzky–golay filter for fnirs
  signal.
\newblock {\em Biocybernetics and Biomedical Engineering}, 39(3):624--637,
  2019.

\bibitem{EEGfeatures}
Poomipat Boonyakitanont, Apiwat Lek-uthai, Krisnachai Chomtho, and Jitkomut
  Songsiri.
\newblock A review of feature extraction and performance evaluation in
  epileptic seizure detection using eeg.
\newblock {\em Biomedical Signal Processing and Control}, 57:101702, 2020.

\bibitem{welch}
Julius~O. Smith.
\newblock {\em Spectral Audio Signal Processing}.
\newblock CCRMA-Stanford University, 2011.
\newblock online book, 2011 edition.

\bibitem{scikit-learn}
F.~Pedregosa, G.~Varoquaux, A.~Gramfort, V.~Michel, B.~Thirion, O.~Grisel,
  M.~Blondel, P.~Prettenhofer, R.~Weiss, V.~Dubourg, J.~Vanderplas, A.~Passos,
  D.~Cournapeau, M.~Brucher, M.~Perrot, and E.~Duchesnay.
\newblock Scikit-learn: Machine learning in {P}ython.
\newblock {\em Journal of Machine Learning Research}, 12:2825--2830, 2011.

\bibitem{Schizo}
Afshin Shoebi, Delaram Sadeghi, Parisa Moridian, Navid Ghassemi, Jonathan
  Heras, Roohallah Alizadehsani, Ali Khadem, Yinan Kong, Saeid Nahavandi,
  Yu-Dong Zhang, and Juan Gorriz.
\newblock Automatic diagnosis of schizophrenia in eeg signals using cnn-lstm
  models.
\newblock {\em Signal Processing}, 2021.

\bibitem{statsmodels}
Skipper Seabold and Josef Perktold.
\newblock statsmodels: Econometric and statistical modeling with python.
\newblock In {\em 9th Python in Science Conference}, 2010.

\bibitem{discussionone}
Miseon Shim, Han-Jeong Hwang, Do-Won Kim, Seung-Hwan Lee, and Chang-Hwan Im.
\newblock Machine-learning-based diagnosis of schizophrenia using combined
  sensor-level and source-level eeg features.
\newblock {\em Schizophrenia Research}, 176(2):314--319, 2016.

\bibitem{discussiontwo}
Omar Odish, Kristinn Johnsen, Paul Someren, Raymund Roos, and J.~Gert~van Dijk.
\newblock Eeg may serve as a biomarker in huntington’s disease using machine
  learning automatic classification.
\newblock {\em Scientific Reports}, 2018.

\end{thebibliography}

\EOD

\end{document}